\documentclass[draft]{amsart}
\usepackage[centertags]{amsmath}
\usepackage{amsfonts}
\usepackage{amssymb}
\usepackage{amsthm}
\usepackage{amsmath}
\usepackage{amscd}
\usepackage{latexsym}
\usepackage{color}
\usepackage{graphics}
\newtheorem{thm}{Theorem}

\newtheorem{lem}{Lemma}
\newtheorem{prop}{Proposition}
\newtheorem{rem}{Remark}
\begin{document}

\title
 {Discrete model of Yang-Mills equations
 in Minkowski space}

\author{ Volodymyr Sushch }

\address{Department of Mathematics, Technical University of Koszalin, Sniadeckich 2,
 75-453 Koszalin, Poland;  Pidstrygach Institute for Applied Problems of Mechanics and
 Mathematics,  Lviv, Ukraine }

\email{sushch@lew.tu.koszalin.pl}



\date{Octovber 2004}
\maketitle

\begin{abstract}
Using methods of differential geometry, a discrete analog
of the Yang-Mills equations in Minkowski space is constructed. The gauge
transformation law in a discrete formulation is given and gauge invariance
of discrete Yang-Mills equations is studied. Difference self-dual and anti-self-dual
equations with respect to the Lorentz metric are presented.
\bigskip
\end{abstract}

\section {Introduction}

The main goal of this paper is to construct a gauge-invariant discrete model
 of Yang-Mills equations in Minkowski space. Based on the formalism described
in \cite{D1} by Dezin, we consider some intrinsically defined geometric discrete model.
A simple two-dimensional discrete model of the classical Yang-Mills equations has
been constructed and studied in \cite{D2}. However, this discrete model is the lacking of
gauge invariance. Some another approaches are proposed in \cite{SM,SD}.
In \cite{SM} a gauge-invariant
discrete analog of the Yang-Mills equations is constructed in Euclidean space
$\Bbb{R}^n$.
We try to define what gauge invariance is in the case of discrete models. The method
described in \cite{SD} is applicable for obtaining a discrete model of the Yang-Mills equations
on the 2-dimensional sphere.

 In this paper we concerned with two problems related to discretization in Min\-kowski
space. First we must determine a combinatorial pseudo-Euclidean space and define a discrete
analog of the Lorentz metric. Note that in this case to define discrete analogs of
the differential and the exterior multiplication can be used the results of \cite{SM}.
As in the continual case these operations do not depend on a metric.
Secondly, given a discrete analog of the connection 1-form, a discrete covariant
derivative must be defined. We always try to be as close to  continual Yang-Mills
theory as possible. Nevertheless, gauge invariance of the discrete Yang-Mills equations
is obtained under some additional conditions (Theorem 1).

 It is known that Yang-Mills theory can be regarded as a non-linear generalization of
Hodge theory in the 4-dimensional case (see \cite{FU}). In Section~5, we construct an operator
 formally adjoint to the discrete covariant differentiation operator. Then we
show how to obtain a discrete analog of the generalized Laplace type operator with respect to
the Lorentz metric.

By analogy with the continual case, one of the questions to be
studied is discrete analogs of the self-dual and anti-self-dual
equations. In Section~6, difference self-dual and anti-self-dual
equations are presented as a system of non-linear matrix
equations.
\section{Preliminaries}
Let $M^4=\Bbb{R}^{1,3}$ be the Minkowski space-time manifold.
Suppose that $M^4$ has the Lorentz metric
$g_{\mu\nu}=diag(-+++)$. Consider the trivial bundle ${P=M^4\times
SU(2)}$. Let $T^\ast P$ be the cotangent bundle of $P$. It is
known (see~\cite{NS}) that a connection can be shown to arise from
a certain 1-form $\omega$ belonging to $T^\ast P$, where $\omega$
is required to have values in the Lie algebra $su(2)$. Let
$(x,g)$, $x\in M^4$, $g\in SU(2)$, be local coordinates of the
bundle $P$. Then $\omega$ is given by $$
\omega=g^{-1}dg+g^{-1}Ag,\eqno (1) $$ where $$
A=\sum_{\alpha,\mu}A_{\mu}^{\alpha}(x)\lambda_{\alpha}
dx^\mu.\eqno (2) $$ Here we take as a basis for $su(2)$ the set
$\{\lambda_\alpha=\frac{\sigma_\alpha}{2i}, \alpha=1,2,3\}$, where
$\sigma_\alpha$  are the standard Pauli matrices. The
$su(2)$-valued 1-form $A$ is called the connection form and the
functions $A_{\mu}^{\alpha}(x)$, connections.

Let the coordinates of $P$ change (locally) from $(x,g)$ to $(x^\prime, g^\prime)$.
Let us only make a change of fibre coordinates, i.e. $x=x^\prime$ and $g^\prime$
is given by
$$
g^\prime=hg, \qquad h\in SU(2).\eqno (3)
$$
So invariance of $\omega$ means that
$$
 g^{-1}dg+g^{-1}Ag=(g^\prime)^{-1}dg^\prime+(g^\prime)^{-1}A^\prime g^\prime.
$$
Under the change of coordinates (3) the invariant 1-form $\omega$ induces a
certain transformation law for the connection form $A$. Taking into account the
fact that $dg^\prime=dhg+hdg$ and $dhh^{-1}+hdh^{-1}=0$, we obtain
$$
A^\prime=hdh^{-1}+hAh^{-1}.\eqno (4)
$$
In Yang-Mills theory this transformation law is called the gauge transformation law.

The curvature 2-form $F$ can be defined as follows
$$
F=dA+A\wedge A. \eqno (5)
$$
We have the tensorial law
$$
F^\prime=hFh^{-1}
$$
 for the change of $F$ under the gauge transformation (4).

Define the covariant exterior differential operator $d_A$ by
$$
d_A\Omega=d\Omega+A\wedge \Omega+(-1)^{r+1}\Omega\wedge A,\eqno (6)
$$
where $\Omega$ is a $su(2)$-valued $r$-form.

Consider the equations
$$
d_A F=0,\eqno (7)
$$
$$
d_A\ast F=0, \eqno (8)
$$
where $\ast$ is the metric adjoint operation (Hodge star). Equations (7), (8)
are called the Yang-Mills equations \cite{FU}. Equation (7) is known as the Bianchi identity.

Let $\Phi, \Psi$ be $su(2)$-valued  $r$-forms on $M^4$. The "inner product" can be defined as
$$
(\Phi, \Psi)=-tr\int_{M^4}\Phi\wedge\ast\Psi,\eqno (9)
$$
where $tr$ is the trace operator. Note that $M^4$ is non-compact. So all forms referring to the
inner product have compact support by assumption.

Then the adjoint operator to $d_A$ can be expressed in the form
$$
\delta_A=\ast^{-1}d_A\ast,
$$
where $\ast^{-1}$ is the inverse operation to $\ast$ ($\ast\ast^{-1}=1$).
Combining the latter with Equation (8) we get
$$
\delta_A F=0.
$$
By virtue of (7), this equation is similar to the criterion for a scalar
differential form to be harmonic \cite{W}.
Thus, if $F$ is a solution of the Yang-Mills equations, then
the following Laplace-Beltrami type equation
$$
(d_A\delta_A+\delta_Ad_A)F=0 \eqno (10)
$$
holds immediately on $M^4$ with respect to the Lorentz metric.

In Minkowski space the self-dual or anti-self-dual equations can be written
as follows
$$
\ast F=\mp iF. \eqno (11)
$$
Since $F$ is $su(2)$-valued, so therefore is $\ast F$, then we must have
${su(2)=isu(2)}$. However, this condition is not
satisfied for $su(2)$ because it is not satisfied for the Lie algebra of any compact
Lie groups \cite{NS}. In this case, to study the instanton problems
 one must choose some non-compact
groups instead $SU(2)$ such as $SL(2,\Bbb{C})$ or $GL(2,\Bbb{C})$ say.

\section{Combinatorial model of Minkowski space}

Following [1], let the tensor product  $C(4)=C\otimes C\otimes C\otimes C$
of a \linebreak  1-dimensional complex be a combinatorial model of Euclidean space
 $\Bbb{R}^4$. The 1-dimensional complex $C$ is defined in the following way.
Let $C^0$ denotes the real linear space of 0-dimensional chains generated by
basis elements $x_\kappa$ (points), $\kappa\in \Bbb{Z}$. It is convenient to
introduce the shift operators  $\tau,\sigma$ in the set of indices by
$$
\tau\kappa=\kappa+1, \qquad \sigma\kappa=\kappa-1.
$$
We denote the open interval $(x_\kappa, x_{\tau\kappa})$ by $e_\kappa$.
One can regard the set $\{e_{\kappa}\}$ as a set of basis elements of
the real linear space $C^1$. Suppose that $C^1$ is the space of 1-dimensional
chains.
Then the 1-dimensional complex (combinatorial real line) is the direct sum
of the introduced spaces $C=C^0\oplus C^1$. The boundary operator $\partial$
in $C$ is given by
$$
\partial x_\kappa=0, \qquad  \partial e_\kappa=x_{\tau\kappa}-x_\kappa.
$$
The definition is extended to arbitrary chains by linearity.

Multiplying the basis elements $x_\kappa, e_\kappa$ in various way we obtain
basis elements of $C(4)$. If $c_p, c_q$ are chains of the indicated dimension,
belonging to the complexes being multiplied, then
$$
\partial(c_p\otimes c_q)=\partial c_p\otimes c_q+(-1)^pc_p\otimes\partial c_q. \eqno (12)
$$
Relation (12) defines the boundary operator in $C(4)$.

We suppose that the combinatorial model of Minkowski space has the same structure
as $C(4)$. We denote only the basis elements corresponding to the time coordinate of
$M^4$ by $\bar x_\kappa$, $\bar e_\kappa$. So, for example, the 1-dimensional basis elements
of $C(4)$ can be written as
$$
e_k^1=\bar e_{k_1}\otimes x_{k_2}\otimes x_{k_3}\otimes x_{k_4}, \qquad
e_k^2=\bar x_{k_1}\otimes e_{k_2}\otimes x_{k_3}\otimes x_{k_4},
$$
$$
e_k^3=\bar x_{k_1}\otimes x_{k_2}\otimes e_{k_3}\otimes x_{k_4},  \qquad
e_k^4=\bar x_{k_1}\otimes x_{k_2}\otimes x_{k_3}\otimes e_{k_4}, \eqno(13)
$$
where $k=(k_1, k_2, k_3, k_4)$ is multiindex, $k_j\in\Bbb{Z}$, $j=1,2,3,4$.

Let us now consider a dual complex to $C(4)$. We define its as the complex of cochains
$K(4)$ with coefficients belonging to $su(2)$ The complex $K(4)$
has a similar structure, namely ${K(4)=K\otimes K\otimes K\otimes K}$, where $K$ is a dual
complex to the 1-dimensional complex $C$. Basis elements of $K$ can be written as
$\{x^\kappa\}, \{e^\kappa\}$. Then an arbitrary basis element of $K(4)$ is given
by ${s^k=\bar s^{k_1}\otimes s^{k_2}\otimes s^{k_3}\otimes s^{k_4}}$, where $s^{k_j}$
is either  $x^{k_j}$ or  $e^{k_j}$.

As in \cite{D2}, we define the pairing operation for arbitrary basis elements
$\varepsilon_k\in C(4)$,  $s^k\in K(4)$ by the rule
$$
<\varepsilon_k, as^k>=\left\{\begin{array}{l}0,\ \varepsilon_k\ne s_k\\
                            a,\ \varepsilon_k=s_k,\ a\in su(2).
                            \end{array}\right. \eqno (14)
$$
The operation (14) is linearly extended to cochains. We will call cochains forms,
emphasizing their relationship with the corresponding continual objects, differential
forms.

The coboundary operator $d^c$ is defined by
$$
<\partial\varepsilon_k, as^k>=<\varepsilon_k, ad^cs^k>. \eqno (15)
$$
The operator $d^c$ is an analog of the exterior differentiation operator.

Let us now introduce in $K(4)$ a multiplication which is an analog
of the exterior multiplication for differential forms. First we
introduce the $r$-dimensional complex $K(r)$, ${r=1,2,3}$,\ in an
obvious notation. Let $s_{(p)}^k$ be an arbitrary $p$-dimensional
basis element of $K(r)$, i.e. the following product \linebreak
${s_{(p)}^k=\bar s^{k_1}\otimes...\otimes s^{k_r}}$ contains
exactly $p$ of the 1-dimensional basis elements $e^{k_{j}}$ and
$r-p$ of the 0-dimensional basis elements  $x^{k_{j}}$,
$k_j\in\Bbb{Z}$, $j=1,...r$. It should be noted that the whole
requisite information about the number and situation of
"components" is contained in the symbol $(p)$. Then, supposing
that the $\cup$-multiplication in $K(r)$ has been defined, we
introduce it for basis elements of $K(r+1)$ by the rule $$
(s^k_{(p)}\otimes s^\kappa)\cup(s^k_{(q)}\otimes s^\mu)=
Q(\kappa,q)(s^k_{(p)}\cup s^k_{(q)})\otimes(s^\kappa\cup s^\mu),
\eqno (16) $$ where $s^k_{(p)}, s^k_{(q)}\in K(r)$,
$s^\kappa(s^\mu)$ is either $x^\kappa(x^\mu)$ or
$e^\kappa(e^\mu)$, $\kappa, \mu\in\Bbb{Z}$, and the signum
function $Q(\kappa, q)$ is equal to $-1$ if the dimension of both
elements $s^\kappa$, $s_{(q)}^k$ is odd and to $+1$ otherwise (see
\cite{D1}). For the basis elements of $K$ the
$\cup$-multiplication is defined as follows $$ x^\kappa\cup
x^\kappa=x^\kappa, \quad e^\kappa\cup x^{\tau\kappa}=e^\kappa,
\quad x^\kappa\cup e^\kappa=e^\kappa, \quad \kappa\in\Bbb{Z}, $$
supposing the product to be zero in all other case. To arbitrary
forms the $\cup$-multiplication can be extended linearly.
Coefficients of forms multiply as matrices.

\begin{prop}{Let $\varphi$ and $\psi$ be arbitrary forms of $K(4)$.
Then $$ d^c(\varphi\cup\psi)=d^c\varphi\cup\psi+(-1)^p\varphi\cup
d^c\psi, \eqno (17) $$ where  $p$ is the dimension of a form
$\varphi$.}
\end{prop}

The proof of Proposition 1 is totally analogous to one in \cite[p.147]{D1} for the case of
discrete forms with real coefficients.

By definition, the coboundary operator $d^c$ and the
$\cup$-multiplication do not depend on a metric. So they have the same structure in
$K(4)$ as in the case of the combinatorial Euclidean space \cite{SM}.
 At the same time, to define a discrete analog of the operation
$\ast$ we must take into account the structure of the Lorentz metric on $K(4)$.
In this case  it is convenient to write the basis elements of the complex $K(4)$
in the form $\bar\mu^\kappa\otimes s^k$, where $s^k$ is a basis element of $K(3)$
and $\bar\mu^\kappa$ is either $\bar x^\kappa$ or $\bar e^\kappa$, $\kappa\in\Bbb{Z}$.

Then we define the operation $\ast$ as follows
$$
\bar\mu^\kappa\otimes s^k\cup\ast(\bar\mu^\kappa\otimes s^k)=
Q(\mu)\bar e^\kappa\otimes e^{k_1}\otimes e^{k_2}\otimes e^{k_3},\eqno (18)
$$
where $Q(\mu)$ is equal to $+1$ if $\bar\mu^\kappa=\bar x^\kappa$ and
to $-1$ if  $\bar\mu^\kappa=\bar e^\kappa$.
\newline
Relation (18) describes the structure of the Lorentz metric in the discrete model.

\section{Discrete Yang-Mills equations}

The discrete analog of the connection 1-form (2) can be written as
$$
A=\sum_{j=1}^4\sum_kA_k^je_j^k, \eqno (19)
$$
where $e_j^k$ is the 1-dimensional basis element of $K(4)$ and $A_k^j\in su(2)$,
\linebreak $k=(k_1,k_2,k_3,k_4), \ k_j\in\Bbb Z$.

Consider the discrete form
$$
h=\sum_kh_kx^k, \eqno (20)
$$
where $x^k$ is the 0-dimensional basis element of $K(4)$ and $h_k\in SU(2)$.
Note that the 0-form (20) does not belong to the complex $K(4)$. But, since \linebreak
$x^k\in K(4)$, the $\cup$-multiplication and the coboundary operator $d^c$ are
generalized on the forms (20) in an obvious way.

Then the discrete analog of the gauge transformation (3), (4) can be written as
$$
g^\prime=h\cup g, \qquad A^\prime=h\cup d^ch^{-1}+h\cup A\cup h^{-1}, \eqno(21)
$$
where $h, h^{-1}, g$ are forms of the type (20). Here we denote by $h^{-1}$ the
form whose coefficients (matrices) are inverse to coefficients of $h$. If $e$
is the 0-form (20) all of whose coefficients are unit elements of the group $SU(2)$,
then we have
$$
h\cup h^{-1}=h^{-1}\cup h=e.
$$
It should be noted that the 0-forms  defined by (20) generate a group by respect to the
$\cup$-multiplication.

Given the discrete analog of the 1-form (1) by the formula
$$
\omega=g^{-1}\cup d^cg+g^{-1}\cup A\cup g,
$$
it is easy to proof that $\omega$ is invariant ($\omega=\omega^\prime$)
under the transformation (21) (see \cite{SM}).

Now consider the 2-form
$$
F=\sum_{j=1}^6\sum_kF_k^j\varepsilon_j^k, \eqno (22)
$$
where $F_k^j\in su(2)$ and $\varepsilon_j^k$ is the 2-dimensional
basis element of $K(4)$. The 2-dimensional basis elements of $K(4)$ can be written
as follows
$$
\varepsilon_1^k=\bar e^{k_1}\otimes e^{k_2}\otimes x^{k_3}\otimes x^{k_4}, \qquad
\varepsilon_2^k=\bar e^{k_1}\otimes x^{k_2}\otimes e^{k_3}\otimes x^{k_4},
$$
$$
\varepsilon_3^k=\bar e^{k_1}\otimes x^{k_2}\otimes x^{k_3}\otimes e^{k_4}, \qquad
\varepsilon_4^k=\bar x^{k_1}\otimes e^{k_2}\otimes e^{k_3}\otimes x^{k_4},
$$
$$
\varepsilon_5^k=\bar x^{k_1}\otimes e^{k_2}\otimes x^{k_3}\otimes e^{k_4}, \qquad
\varepsilon_6^k=\bar x^{k_1}\otimes x^{k_2}\otimes e^{k_3}\otimes e^{k_4},
$$
where $k_i\in\Bbb{Z}$, $i=1,2,3,4.$

We define the discrete analog of the curvature 2-form by the formula
$$
F=d^cA+A\cup A. \eqno (23)
$$

\begin{prop}  Under the gauge transformation (21) the curvature form
(23) changes as $$ F^\prime=h\cup F\cup h^{-1}. $$
\end{prop}
 \begin{proof} The proof closely follows the proof Theorem 2 of
\cite{SM}. Using (21) and (17) we compute $$ d^c A^\prime=d^ch\cup
d^ch^{-1}+d^ch\cup A\cup h^{-1}+ h\cup d^cA\cup h^{-1}-h\cup A\cup
d^ch^{-1}. $$ Since $d^ce=0$ by definition of $d^c$, we have $$
d^c(h\cup h^{-1})=d^ch\cup h^{-1}+h\cup d^ch^{-1}=0 $$ and so $$
d^ch\cup h^{-1}=-h\cup d^ch^{-1}.\eqno (24) $$ Taking into account
(24), we obtain $$
\begin{array}{l}
A^\prime\cup A^\prime=(h\cup d^ch^{-1}+h\cup A\cup h^{-1})\cup
(h\cup d^ch^{-1}+h\cup A\cup h^{-1})\\
 =-d^ch\cup d^ch^{-1}+h\cup A\cup d^ch^{-1}-d^ch\cup A\cup h^{-1}+
h\cup A\cup A\cup h^{-1}.
\end{array}
$$
Then we finally have
$$
\begin{array}{l}
F^\prime=d^cA^\prime+A^\prime\cup A^\prime=h\cup d^c A\cup h^{-1}
+h\cup A\cup A\cup h^{-1}\\
\quad {}=h\cup(d^c A+A\cup A)\cup h^{-1}.
\end{array}
$$
\end{proof}

From the definition (23) one easily derives that the curvature form $F$
satisfies the identity
$$
d^cF+A\cup F-F\cup A=0. \eqno (25)
$$
The comparison of (25) with (7) yields a discrete analog of the Bianchi identity. Define now
the discrete analog of the exterior covariant differentiation operator by setting
$$
d_A^c\Omega=d^c\Omega+A\cup\Omega+(-1)^{r+1}\Omega\cup A,
$$
where $\Omega$ is an arbitrary $r$-form of $K(4)$.
Then Identity (25) can be rewritten as
$$
d^c_AF=0.
$$
In similar manner, we obtain the discrete analog of Equation (8)
$$
d_A^c\ast F\equiv d^c\ast F+A\cup\ast F-\ast F\cup A=0. \eqno (26)
$$

Let $\tau_{ij}$  ($\sigma_{ij}$),\  $i,j=1,2,3,4,\  i\ne j$, be the shift operator
acting as the operator $\tau$ ($\sigma$) by the $i$-th, $j$-th components of the
multiindex $k=(k_1,k_2,k_3,k_4)$. For example,
$$
\tau_{12}k=(\tau k_1,\tau k_2,k_3,k_4),\qquad
 \sigma_{23}k=(k_1,\sigma k_2,\sigma k_3,k_4).
$$
Using the definition (18) we compute
$$
\ast F=\sum_k (F_{\sigma_{34}k}^6\varepsilon_1^k-
F_{\sigma_{24}k}^5\varepsilon_2^k +
F_{\sigma_{23}k}^4\varepsilon_3^k
- F_{\sigma_{14}k}^3\varepsilon_4^k+
F_{\sigma_{13}k}^2\varepsilon_5^k -
F_{\sigma_{12}k}^1\varepsilon_6^k).
\eqno (27)
$$

\begin{lem} Let $h$ be a discrete 0-form. Then we have $$ \ast(h\cup
f)=h\cup\ast f \eqno (28) $$ for an arbitrary $p$-form $f\in
K(4)$.
\end{lem}

\begin{proof}
 Any $p$-form $f\in K(4)$ can be expressed as
$$
f=\sum_k f_k^{(p)}s_{(p)}^k,
$$
where  $f_k^{(p)}\in su(2)$ and $s_{(p)}^k$ is the $p$-dimensional basis element of $K(4)$.
By definition, we have  $x^k\cup s_{(p)}^k=s_{(p)}^k$ for an arbitrary 0-dimensional
 basis element $x^k$ of $K(4)$.
Hence,
$$
h\cup f=\left(\sum_k h_kx^k\right)\cup\left(\sum_k f_k^{(p)}s^k_{(p)}\right)
=\sum_k h_kf_k^{(p)}s^k_{(p)}.
$$
Then we obtain
$$
\ast(h\cup f)=\sum_k h_kf_k^{(p)}\ast s^k_{(p)}=h\cup\sum_k f_k^{(p)}\ast s^k_{(p)}=
h\cup\ast f.
$$
\end{proof}

\begin{lem} We have
$$ \ast(f\cup h)=\ast f\cup h \eqno (29) $$ for an arbitrary
2-form $f\in K(4)$ if and only if coefficients of a 0-form $h$
satisfy the following conditions $$ h_k=h_{\sigma_{ij}k} \eqno
(30) $$ for all $i,j=1,2,3,4, i\ne j$.
\end{lem}

\begin{proof} From the definition (16) one easily derives that
in the form $f\cup h$ two indices in the coefficients $h_k$ are
shifted and we have $$ f\cup
h=\sum_k(f_k^1h_{\tau_{12}k}\varepsilon_1^k+f_k^2h_{\tau_{13}k}\varepsilon_2^k+
f_k^3h_{\tau_{14}k}\varepsilon_3^k+f_k^4h_{\tau_{23}k}\varepsilon_4^k+
f_k^5h_{\tau_{24}k}\varepsilon_5^k+f_k^6h_{\tau_{34}k}\varepsilon_6^k).
$$ Since $$ \ast\varepsilon_1^k=-\varepsilon_6^{\tau_{12}k},
\qquad \ast\varepsilon_2^k=\varepsilon_5^{\tau_{13}k}, \qquad
\ast\varepsilon_3^k=-\varepsilon_4^{\tau_{14}k}, $$ $$
\ast\varepsilon_4^k=\varepsilon_3^{\tau_{23}k}, \qquad
\ast\varepsilon_5^k=-\varepsilon_2^{\tau_{24}k}, \qquad
\ast\varepsilon_6^k=\varepsilon_1^{\tau_{34}k}, $$ we obtain $$
\ast(f\cup
h)=\sum_k(-f_k^1h_{\tau_{12}k}\varepsilon_6^{\tau_{12}k}
+f_k^2h_{\tau_{13}k}\varepsilon_5^{\tau_{13}k}-
f_k^3h_{\tau_{14}k}\varepsilon_4^{\tau_{14}k} $$ $$
\qquad\qquad\qquad +f_k^4h_{\tau_{23}k}\varepsilon_3^{\tau_{23}k}-
f_k^5h_{\tau_{24}k}\varepsilon_2^{\tau_{24}k}+
f_k^6h_{\tau_{34}k}\varepsilon_1^{\tau_{34}k}).\eqno (31) $$
Taking into account the relation $$ \sum_kh_kx^k=\sum_kh_{\tau
k}x^{\tau k}, $$ we compute $$ \ast f\cup h=\sum_k(-f_k^1h_{\tau
k}\varepsilon_6^{\tau_{12}k}+ f_k^2h_{\tau
k}\varepsilon_5^{\tau_{13}k}- f_k^3h_{\tau
k}\varepsilon_4^{\tau_{14}k} $$ $$ \qquad\qquad\qquad
+f_k^4h_{\tau k}\varepsilon_3^{\tau_{23}k}- f_k^5h_{\tau
k}\varepsilon_2^{\tau_{24}k}+ f_k^6h_{\tau
k}\varepsilon_1^{\tau_{34}k}),\eqno (32) $$ where $\tau k=(\tau
k_1,\tau k_2,\tau k_3,\tau k_4)$.

Inserting (31), (32) into (29), we get
$$
h_{\tau k}=h_{\tau_{12}k}=h_{\tau_{13}k}=h_{\tau_{14}k}=h_{\tau_{23}k}
=h_{\tau_{24}k}=h_{\tau_{34}k}
$$
for an arbitrary $k$. Clearly, these relations imply (30).

On other hand,  Conditions (30) we can rewritten as follows
$h_{\tau k}=h_{\tau_{ij}k}$ for all $i,j=1,2,3,4,\  i\ne j$.
Substituting the latter into (31) and comparing (31) and (32), we
obtain (29).

\end{proof}

It should be noted that in the Lemmas we can taken the 0-form $h$
either as an element of $K(4)$ or as a form of the type (20).

Conditions (30) mean that the "diagonal components" of the 0-form $h$
are equal in all plans as shown in Fig.1. Remind that we regard $h$ as
the function over the points $x_k$ and ${h|_{x_k}=<x_k, h>=h_k}$.

\begin{picture}(300,180)
\put(89,40){.}
\put(89,90){.}
\put(139,40){.}
\put(139,140){.}
\put(189,90){.}
\put(139,90){.}
\put(189,140){.}
\put(90,40){\line(0,1){60}}
\put(90,40){\line(1,0){60}}
\put(90,40){\line(0,-1){10}}
\put(90,40){\line(-1,0){10}}
\put(90,90){\line(1,0){110}}
\put(140,40){\line(0,1){110}}
\put(90,90){\line(-1,0){10}}
\put(140,40){\line(0,-1){10}}
\put(140,140){\line(1,0){60}}
\put(190,90){\line(0,1){60}}
\put(140,140){\line(-1,0){10}}
\put(190,90){\line(0,-1){10}}
\put(143,95){$h_k$}
\put(193,95){$h_{\tau_ik}$}
\put(193,145){$h_{\tau_{ij}k}$}
\put(143,145){$h_{\tau_jk}$}
\put(143,45){$h_{\sigma_jk}$}
\put(93,45){$h_{\sigma_{ij}k}$}
\put(93,95){$h_{\sigma_ik}$}
\put(110,5){Fig. 1. (in the plan ($k_i,k_j$)).}

\end{picture}
\begin{prop} The set of 0-forms (20) satisfying  Conditions (30) is a
group under $\cup$-multiplication.
\end{prop}
\begin{proof}
The claim is obvious. By the definition (16), the product of any
0-forms is a 0-form and indices in coefficients do not shift. From this the
result follows at once.
\end{proof}

\begin{thm} Under Conditions (30) the Yang-Mills
equation (26) is gauge invariant.
\end{thm}

Here gauge invariance is understood as follows. If $A(F)$ is a solution of
Equation (26), then  $A^\prime(F^\prime)$ is also a solution of (26).

\begin{proof}By Proposition 3, the form $h^{-1}$ satisfies  Conditions (30).
Using Proposition 2 from Lemma 1 and 2 we have
$$
\ast F^\prime=h\cup\ast F\cup h^{-1}.
$$
Now express $d^c_{A^\prime}\ast F^\prime$ in terms of $F, A$. Applying (17) we
compute
$$
d^c\ast F^\prime=d^ch\cup\ast F\cup h^{-1}+h\cup d^c\ast F\cup h^{-1}+
h\cup\ast F\cup d^ch^{-1}.
$$
Taking into account (21) and (24), we obtain
$$
A^\prime\cup\ast F^\prime=-d^ch\cup\ast F\cup h^{-1}+
h\cup A\cup\ast F\cup h^{-1}
$$
and
$$
\ast F^\prime\cup A^\prime=h\cup\ast F\cup d^ch^{-1}+
h\cup\ast F\cup A\cup h^{-1}.
$$
Thus,
$$
d^c_{A^\prime}\ast F^\prime=h\cup d_A^c\ast F\cup h^{-1}.
$$

\end{proof}

\section{The operator formally adjoint to $d_A^c$}

Let $V\subset C(4)$ be some fixed "domain" of the complex $C(4)$.
We can written $V$ as follows
$$
V=\sum_kV_k, \qquad k=(k_1,k_2,k_3,k_4),\quad k_i=1,2, ...,N_i, \eqno (33)
$$
where $V_k=\bar e_{k_1}\otimes e_{k_2}\otimes e_{k_3}\otimes e_{k_4}$
is the 4-dimensional basis element of $C(4)$. We agree that
in what follows the subscripts $k_i,\  i=1,2,3,4$, always run the set
of values indicated in (33).
In this section we suppose that coefficients of the discrete forms are
vanished on $C(4)\setminus V$. Then the "inner product" for forms
$\Phi, \Psi\in K(4)$ of the same degree is defined by the relation
$$
(\Phi, \Psi)_V=-tr<V, \Phi\cup\ast\Psi>. \eqno (34)
$$
For the forms of different degrees the product (34) is set equal to zero.

The definition imitates correctly the continual case (Relation (9)). It follows
from (18) that for the basis elements $e_i^k$ and $\varepsilon_j^k$ we have
$e_1^k\cup\ast e_1^k=-V^k$,\ $e_i^k\cup\ast e_i^k=V^k$\ for $i=2,3,4$, \
$\varepsilon_j^k\cup\ast\varepsilon_j^k=-V^k$ \ for $j=1,2,3$ and
$\varepsilon_j^k\cup\ast\varepsilon_j^k=V^k$ \ for $j=4,5,6$. Then we obtain
$$
(A, A)_V=-tr\sum_k\left[-(A_k^1)^2+(A_k^2)^2+(A_k^3)^2+(A_k^4)^2\right]
$$
and
$$
(F, F)_V=-tr\sum_k\left[-(F_k^1)^2-(F_k^2)^2-(F_k^3)^2+(F_k^4)^2+(F_k^5)^2+
(F_k^6)^2\right],
$$
where $A$ is an 1-form (19) and $F$ is a 2-form (22).

\begin{prop} Let $\Phi\in K(4)$ be an 1-form and $\Psi\in K(4)$ be a
2-form. Then we have $$ (d^c\Phi, \Psi)_V=(\Phi, \delta^c\Psi)_V,
$$ where $$ \delta^c\Psi=\ast^{-1}d^c\ast\Psi \eqno(35) $$ is the
operator formally adjoint to $d^c$.
\end{prop}
\begin{proof} From (17) and (34) we obtain $$\begin{array}{rcl} (d^c\Phi,
\Psi)_V&=&-tr<V, d^c\Phi\cup\ast\Psi)>\\ &=&-tr<V,
d^c(\Phi\cup\ast\Psi)>-tr<V, \Phi\cup d^c\ast\Psi>\\
&=&-tr<\partial V, \Phi\cup\ast\Psi>-
tr<V,\Phi\cup\ast(\ast^{-1}d^c\ast\Psi)>\\ &=&-tr<\partial V,
\Phi\cup\ast\Psi>+ (\Phi,\ \ast^{-1}d^c\ast\Psi)_V,
\end{array}
$$
where we used $\ast\ast^{-1}=1$.

\noindent
Let $\tilde e_k^j, \ j=1,2,3,4,$ denote the 3-dimensional basis element
of $C(4)$. Using (12) we derive that
$$
\partial V=\sum_k (\tilde e_{\tau N_1,k_2,k_3,k_4}^1-
\tilde e_{1,k_2,k_3,k_4}^1-\tilde e_{k_1,\tau N_2,k_3,k_4}^2+
\tilde e_{k_1,1,k_3,k_4}^2
$$
$$
\qquad\qquad +\tilde e_{k_1,k_2,\tau N_3,k_4}^3-
\tilde e_{k_1,k_2,1,k_4}^3-\tilde e_{k_1,k_2,k_3,\tau N_4}^4+
\tilde e_{k_1,k_2,k_3,1}^4),
$$
where
$$
\tilde e_k^1=\bar x_{k_1}\otimes e_{k_2}\otimes e_{k_3}\otimes e_{k_4}, \qquad
\tilde e_k^2=\bar e_{k_1}\otimes x_{k_2}\otimes e_{k_3}\otimes e_{k_4},
$$
$$
\tilde e_k^3=\bar e_{k_1}\otimes e_{k_2}\otimes x_{k_3}\otimes e_{k_4}, \qquad
\tilde e_k^4=\bar e_{k_1}\otimes e_{k_2}\otimes e_{k_3}\otimes x_{k_4}.
$$
Computing the "boundary components" of the form $\Phi\cup\ast\Psi$ we obtain
the linear combination of the following products:
$$
\Phi_{k_1...\tau N_i...k_4}^j\cdot\Psi_{k_1...N_i...k_4}^r \
\mbox{and}\
\Phi_{k_1...0...k_4}^j\cdot\Psi_{k_1...0...k_4}^r,\ i,j=1,2,3,4,\
 r=1,2,...,6.
$$
Since we have ${\Phi_{k_1...\tau N_i...k_4}^j=\Psi_{k_1...0...k_4}^r=0}$
for all $i,j,r$ by assumption, it follows that
 ${<\partial V,\Phi\cup\ast\Psi>=0}$.

\end{proof}

For the 2-form $F$ using (27) and the definition of $d^c$ we can rewritten
(35) in the form
$$\begin{array}{r}
\delta^cF=\sum_k[(\Delta_{k_2}F_{\sigma_2k}^1+
\Delta_{k_3}F_{\sigma_3k}^2+\Delta_{k_4}F_{\sigma_4k}^3)e_1^k\\
+(\Delta_{k_1}F_{\sigma_1k}^1+\Delta_{k_3}F_{\sigma_3k}^4+
\Delta_{k_4}F_{\sigma_4k}^5)e_2^k\\
+(\Delta_{k_1}F_{\sigma_1k}^2-\Delta_{k_2}F_{\sigma_2k}^4+
\Delta_{k_4}F_{\sigma_4k}^6)e_3^k\\
+(\Delta_{k_1}F_{\sigma_1k}^3-\Delta_{k_2}F_{\sigma_2k}^5-
\Delta_{k_3}F_{\sigma_3k}^6)e_4^k].
\end{array}
$$
Here we denote by $\Delta_{k_i}F_k^j$ the difference
$F_{\tau_ik}^j-F_k^j,\ j=1,2,...,6,$ and \linebreak
$\tau_ik={k_1...\tau k_i...k_4}$,\  $\sigma_ik={k_1...\sigma k_i...k_4}$, \
$i=1,2,3,4$.

\begin{lem} For any 1-form $\Phi\in K(4)$ and 3-form $\Psi\in K(4)$ the
following relation holds $$ tr<V, \Phi\cup\Psi>=-tr<V,
\Psi\cup\ast\ast\Phi>.\eqno(36) $$
\end{lem}
\begin{proof}The forms $\Phi$ and $\Psi$ can be expressed as $$
\Phi=\sum_{i=1}^4\sum_k\Phi_k^ie_i^k, \qquad
\Psi=\sum_{i=1}^4\sum_k\Psi_k^i\tilde e_i^k, $$ where $\Phi_k^i,\
\Psi_k^i\in su(2)$ and $\tilde e_i^k$ is the 3-dimensional basis
element of $K(4)$.

Using (14), (16) we compute
$$
tr<V, \Phi\cup\Psi>=tr\sum_k(\Phi_{\sigma_1k}^1\cdot\Psi_k^1-
\Phi_{\sigma_2k}^2\cdot\Psi_k^2+\Phi_{\sigma_3k}^3\cdot\Psi_k^3
-\Phi_{\sigma_4k}^4\cdot\Psi_k^4),
$$
where $\sigma_ik=(k_1...\sigma k_i...k_4)$.

On the other hand, since
$$
\ast\ast\Phi=\sum_{i=1}^4\sum_k\Phi_k^ie_i^{\tau k}, \quad
\tau k=(\tau k_1,\tau k_2,\tau k_3,\tau k_4),\eqno (37)
$$
we have
$$
tr<V, \Psi\cup\ast\ast\Phi>=tr\sum_k(-\Psi_k^1\cdot\Phi_{\sigma_1k}^1+
\Psi_k^2\cdot\Phi_{\sigma_2k}^2-\Psi_k^3\cdot
\Phi_{\sigma_3k}^3+\Psi_k^4\cdot\Phi_{\sigma_4k}^4)
$$
$$
\qquad\qquad\qquad\qquad=tr\sum_k(-\Phi_{\sigma_1k}^1\cdot
\Psi_k^1+\Phi_{\sigma_2k}^2\cdot\Psi_k^2-\Phi_{\sigma_3k}^3\cdot
\Psi_k^3+\Phi_{\sigma_4k}^4\cdot\Psi_k^4),
$$
where we used $tr(\Phi_k^i\cdot\Psi_k^i)=tr(\Psi_k^i\cdot\Phi_k^i)$.
From this the result follows at once.

\end{proof}

It should be noted that in the continual case we have the equality
$$
tr(\varphi\wedge\psi)=(-1)^{pq}tr(\psi\wedge\varphi),
$$
 where $\varphi$ and
$\psi$ are matrix-valued differential forms of degree $p,q$, respectively.
Unfortunately, this equality has not an exact analog in our formalism.

\begin{thm}For any 2-form $F\in K(4)$ the formal adjoint operator to
$d_A^c$ acts as follows $$ \delta_A^cF=\ast^{-1}(d^c\ast F-\ast
F\cup\ast\ast A+A\cup\ast F).\eqno (38) $$
\end{thm}
\begin{proof} We will compute the operator $\delta_A^c$ defined by the
relation $$ (d_A^c\Phi, F)_V=(\Phi, \delta_A^cF)_V, $$ where
$\Phi$ is an 1-form.

Using (36) and (34) we have
$$\begin{array}{l}
 (d_A^c\Phi,\ F)_V=-tr<V,\ d_A^c\Phi\cup\ast F>\\=-tr<V,\ d^c\Phi\cup\ast F>
-tr<V,\ (A\cup\Phi\cup\ast F+ \Phi\cup A\cup\ast F)>\\
=-tr<V,\ d^c\Phi\cup\ast F>+tr<V,\ (\Phi\cup\ast F\cup\ast\ast A-
\Phi\cup A\cup\ast F)>\\
=(d^c\Phi,\ F)_V+tr<V,\ (\Phi\cup\ast\ast^{-1}(\ast F\cup\ast\ast A)-
\Phi\cup\ast\ast^{-1}(A\cup\ast F))>\\
=(\Phi,\ \delta^c F)_V-(\Phi,\ \ast^{-1}(\ast F\cup\ast\ast A))_V+
(\Phi,\ \ast^{-1}(A\cup\ast F))_V\\
=(\Phi,\ \delta^cF-\ast^{-1}(\ast F\cup\ast\ast A)+\ast^{-1}(A\cup\ast F))_V.
\end{array}
$$

\end{proof}

In the continual case, if we choose the Lorentz metric, then
$$
\ast\ast A=A
$$
for an arbitrary differential 1-form $A$.

\noindent
Hence the Yang-Mills equation (8) can be rewritten as follows
$$
 d\ast F+A\wedge\ast F-\ast F\wedge\ast\ast A=0.\eqno (39)
$$
It follows that a discrete analog of Equation (39) (or (8)) can be given by
$$
d_A^c\ast F\equiv d^c\ast F+A\cup\ast F-\ast F\cup\ast\ast A=0.\eqno (40)
$$
Comparing the latter and (38) we obtain
$$
\delta_A^cF=\ast^{-1}d_A^c\ast F.
$$
Thus, if the discrete curvature 2-form $F$ is a solution of Equation (40), then
the Laplace type  equation
$$
(d_A^c\delta_A^c+\delta_A^cd_A^c)F=0
$$
holds immediately. This equation we call a discrete analog of Equation (10).

It should be noted that in our discrete model the operation $\ast^2=\ast\ast$
is equivalent to a shift with corresponding sign (see (37)). So, unfortunately,
Equation (40) differs from Equation (26). The possibility of involute
$(\ast\ast=1)$  definition of $\ast$ is discussed in \cite{SM2}.

\section{ Discrete models of the self-dual and anti-self-dual
equations}

In this section we will construct a difference analog of Equations (9).
For this reason we take the group $SL(2,\Bbb C)$ instead $SU(2)$. Let the
components $F_k^j$  of the curvature form $F$ be belonging to $sl(2,\Bbb C)$.
Combining (22) with (27) the discrete self-dual equation $\ast F=iF$ can be
written as follows
$$
F_{\sigma_{34}k}^6=iF_k^1, \qquad -F_{\sigma_{24}k}^5=iF_k^2, \qquad
F_{\sigma_{23}k}^4=iF_k^3,
$$
$$
-F_{\sigma_{14}k}^3=iF_k^4, \qquad F_{\sigma_{13}k}^2=iF_k^5, \qquad
-F_{\sigma_{12}k}^1=iF_k^6
$$
for all $k=(k_1,k_2,k_3,k_4), \ k_r\in\Bbb Z, \ r=1,2,3,4$.
From the latter we obtain
$$
F_{\sigma k}^6=iF_{\sigma_{12}k}^1=-i^2F_k^6=F_k^6, \qquad
F_{\sigma k}^5=-iF_{\sigma_{13}k}^2=-i^2F_k^5=F_k^5
$$
and similarly for any other components $F_k^j, \ j=1,2,...,6$.

So we have
$$
F_k^j=F_{\sigma k}^j. \eqno (41)
$$
We call Equations (41)  difference self-dual equations.

In similar manner, we obtain the difference anti-self-dual equations
$$
F_k^j=-F_{\sigma_k}^j. \eqno (42)
$$

\begin{prop} For any 2-form $F$ such that $F_k^j=\pm F_{\sigma k}^j$ we
have $$ \ast\ast F=\mp F. $$
\end{prop}
\begin{proof}
$$
\ast\ast F=\ast(\pm iF)=\pm i\ast F=\pm i^2F=\mp F.
$$

\end{proof}

Using (23) Equations (41) can be rewritten as
$$
\Delta_{k_i}A_k^r-\Delta_{k_r}A_k^i+A_k^i\cdot A_{\tau_ik}^r
-A_k^r\cdot A_{\tau_rk}^i=
\Delta_{k_i}A_{\sigma k}^r-\Delta_{k_r}A_{\sigma k}^i+
A_{\sigma k}^i\cdot A_{\sigma\tau_ik}^r
-A_{\sigma k}^r\cdot A_{\sigma\tau_rk}^i,
$$
where $\sigma\tau_ik=(\sigma k_1...k_i...\sigma k_4)$, \
 $i<r, \ r=1,2,3,4$ and $A_k^r\in sl(2,\Bbb C)$ is a component of
the connection form $A$. Similarly, Equations (42) become
$$
\Delta_{k_i}A_k^r-\Delta_{k_r}A_k^i+A_k^i\cdot A_{\tau_ik}^r
-A_k^r\cdot A_{\tau_rk}^i=\Delta_{k_r}A_{\sigma k}^i
-\Delta_{k_i}A_{\sigma k}^r-
A_{\sigma k}^i\cdot A_{\sigma\tau_ik}^r
+A_{\sigma k}^r\cdot A_{\sigma\tau_rk}^i.
$$

\begin{prop} Let $F\in K(4)$ be a 2-form with compact support. Then the
discrete self-dual (anti-self-dual) equations have the unique
solution $F=0$.
\end{prop}

\begin{proof} Since Equations (41) (Equations (42)) hold for all
$k=(k_1,k_2,k_3,k_4)$, $k_r\in\Bbb Z$, then the assertion is
obvious.

\end{proof}
\begin{rem}In the continual case we must write the self-dual and anti-self-dual
equations in the form (11) because we have $\ast\ast F=-F$ for the
Lorentz metric. In the case of the discrete model it is easy to
check that in $K(4)$ we have $$ \ast\ast F=-\sum_{j=1}^6\sum_k
F_{\sigma k}^j\varepsilon_j^k. $$ It should now be clear that a
discrete model of Equation (11) can be defined as follows $\ast
F=\pm F$. Then we obtain the following difference equations
$$
F_k^j=\mp F_{\sigma k}^j $$ for all $j=1,2, ...,6, \
k=(k_1,k_2,k_3,k_4), \ k_r\in \Bbb Z$.
Therefore, on opposite to
the continual case we can study these equations for the group
$SU(2)$, i. e. for $F$ with  components $F_k^j\in su(2)$.
\end{rem}

\end{document}